\documentclass[review]{elsarticle}
\usepackage{lineno,hyperref}
\journal{Journal of \LaTeX\ Templates}
\usepackage[russian,english]{babel}

\usepackage{amsthm}
\theoremstyle{plain}

\newtheorem{theorem}{Theorem}
\newtheorem{definition}{Definition}
\newtheorem{lemma}{Lemma}
\newtheorem{corollary}{Corollary}
\newtheorem*{proof*}{Proof}
\usepackage{lineno,hyperref}
\usepackage{ucs} 
\usepackage[russian]{babel}  
\usepackage{tikz} 
\begin{document}
	
	\begin{frontmatter}
		
		\title{ New global estimations of the Cauchy problem for the Navier-Stokes equations .}
		
		\author{Durmagambetov A.A\fnref{myfootnote}}
		\address{}
		\fntext[myfootnote]{}
		\ead[url]{}
		
		\ead{aset.durmagambet@gmail.com}
		
		\address[mymainaddress]{010000, Kazakhstan }
		
		\begin{abstract}
		 The paper presents results, indicating that embedding theorems do not allow to study a process of a catastrophe formation. In fact, the paper justifies Terence Tao’s pessimism about a failure of modern mathematics to solve the Navier-Stokes problem.  An alternative method is proposed for dealing with the gradient catastrophe by studying Fourier transformation for a function and selecting a function singularity through phase singularities of Fourier transformation for a given function.
			The analytic properties of the scattering amplitude are discussed in ${R^{3}}$, and a representation of the potential is obtained using the scattering amplitude. A uniform time estimation of the Cauchy problem solution for the Navier-Stokes equations is provided.
		\end{abstract}
		
		\begin{keyword}
			{Schr\"{o}dinger's equation; potential; scattering amplitude; Cauchy problem; Navier--Stokes equations; Fourier transform, the global solvability and uniqueness of the Cauchy problem,the loss of smoothness,The Millennium Prize Problems}
			\MSC[2010] 11M26 {AMS 11}
		\end{keyword}
		
	\end{frontmatter}

	\section {INTRODUCTION}
	The research presents a process of gradient catastrophe formation under conditions of phase change. The paper shows that classical methods of the function estimation theory in context of Sobolev- Schwartz Space Theory are not suitable for studying gradient catastrophe problem. Results presented here, show that the embedding theorems do not allow to study a process of a catastrophe formation. Actually, the paper justifies Terence Tao’s pessimism about a failure of using present mathematical methods for solving the Navier-Stokes problem.  An alternative method is proposed for studying gradient catastrophe by applying Fourier transformation to a function and selecting function singularity through phase singularities of Fourier transformation for a given function. We know a general definition of a gradient catastrophe - an unbounded increase of a function derivative upon conditions of boundedness of the function itself. This phenomenon occurs in various problems of hydrodynamics, such as a formation of shock waves, weather fronts, hydraulic and seismic fracturing, and others. In modern physics and mathematics, as well as in many other areas of science and technology, this phenomenon is considered as a very difficult problem, both from a theoretical and applied perspective.  From a theoretical point of view this is important as we have to know how to describe qualitative changes in processes, which are manifested in appearance of new quality objects during a process of description model evolution, and in the context of applied research, the problem is facing numerical instability in the event of a gradient catastrophe formation. Thus, we approach an important obstacle while using modeling - a barrier created by the gradient catastrophe. Since, on the one hand, the gradient catastrophe is still unknown phenomenon,  it is very important from a practical point of view, because the phenomenon is connected with the most interesting and important aspects of reality. Terence Tao formulated and illustrated this in [1] based on the Millennium problem stated by Clay Institute for the Navier-Stokes equations. Our point of view on these issues agrees with one, stated in article [1]-[7] but in our research we propose a way to solving these problems.
	Our point of view is that the modern mathematical methods of the theory of functions dedicated to the function estimation have ignored such an important component of the Fourier transformation as its phase.  Our research is outlined as follows:  first, we give examples of the gradient catastrophe caused by the phase change, and then proceed to an expansion of classes of functions subjected to the gradient catastrophe. Our final results lie in the nonlinear representation of functions showing some new classification of functions through a phase classification. In addition, the notions of discreteness and continuity of functions are naturally merged. And, in our opinon,  this leads to understanding of how discrete objects are born under a continuous change of the world. Discrete objects are associated with discrete spectrum of the Liouville- Schrödinger equations. And they, as it is known, reflect the wave nature of things.  But here, we abstract away from the quantum formalism, because our goal lies in a purely mathematical approach to the analysis of the arbitrary functions. For the analysis of which, we formally consider a function as a potential of the Schrödinger equation.  At the same time we come across the concepts that generated by the Liouville- Schrödinger equations. These concepts allow to classify and estimate functions by a phase generated by discrete spectrum of the Liouville equation.
	
	\section{Results for the one-dimensional case}
	Let us consider one-dimensional function $ {f}  $ and its Fourier transformation  $  \tilde{f} $. Using  notions of module and phase, we write Fourier transformation in the following form $  \tilde{f}=|\tilde{f}|\exp(i\Psi) $ , where $ \Psi $ is phase.  To cite  Plancherel equality: $ ||f||_{L_{2}}=Const||\tilde{f} ||_ {L_{2}} $. Here we can see that a phase is not contributed to determination of X norm. To estimate a maximum we have a simple estimate as  $  max|f|^2 \le 2||f||_{L_{2}} ||\nabla f||_{L_{2}}   $.Now we have an estimate of the function maximum in which a phase is not involved. Let us consider a behavior of a progressing wave running with a constant velocity of $  v=a$ described by function $ {F(x,t)=f(x+at)}  $.  For its Fourier transformation along x variable we have  $  \tilde{F}=\tilde{f}exp(iatk) $.  Again in this case we can see that when we will be studying a module of the Fourier transformation, we will not obtain major physical information about the wave, such as its velocity and location of the wave crest because of $ |\tilde{F}|=|\tilde{f}| $ . These two examples show 	w  eaknesses of studying Fourier transformation.  On the other hand, many researchers focus on the study of functions using embedding theorem, but in the embedding theorems main object of the study is module of function. But as we have seen in given examples, a phase is a main physical characteristic of a process, and as we can see in the mathematical studies, which use embedding theorems with energy estimates, the phase disappears. Along with phase, all reasonable information about physical process disappears, as demonstrated by Terence Tao [1] and other research considerations.  In fact, he built progressing waves that are not followed energy estimates. Let us proceed with more essential analysis of influence of the phase on behavior of functions.

	\begin  {theorem}
	There are functions of $ W_{2}^{1}(R)$ with a constant rate of the norm for a gradient catastrophe of which a phase change of its Fourier transformation is sufficient.
	\end  {theorem}
	Proof: To prove this, we consider a sequence of testing functions $  \tilde{f_{n}}=\Delta/(1+k^2),\Delta=(i-k)^n/(i+k)^n $. it is obvious  that  $|\tilde{f_{n}}|=1/(1+k^2).$
	$  max|f_{n}|^2 \le 2||f_{n}||_{L_{2}} ||\nabla f_{n}||_{L_{2}} \le Const   $.. Calculating the Fourier transformation of these testing functions, we obtain: 
\begin{equation}
		f_{n}(x) =	x(-1)^{(n-1)}2\pi \exp(-x)L^1_{(n-1)}(2x)  if\,\, x> 0 ,\,\,
	f_{n}(x) =		0  \,\,\,if \,\,\, x\le 0 
	\end{equation}
	where  $ L^1_{(n-1)}(2x)$ is a Laguerre polynomial
	
	. Now we see that the functions are equibounded and derivatives of these functions will grow with the growth of $ {n}.  $
	Thus, we have built an example of a sequence of the bounded functions of  $ W_{2}^{1}(R)$ which have a constant norm  $ W_{2}^{1}(R)$  and this sequence converges to a discontinuous function.

\section{Results for the three-dimensional case }
Consider Schrodinger'sequation:
\begin{equation}
\label {eq:1}
-\Delta_{x}\Psi  +q\Psi  =k^{2}\Psi, ~k\in C  
\end{equation}
Let $ \Psi_{+ } (k,\theta,x)  $ be a solution of ( \ref{eq:1} )with the following asympotic behavior:
\begin{equation}\label{eq:2}
\Psi_{+ } (k,\theta,x)=\Psi _{0}(k,    
\theta, x)+\frac{e^{ ik|x|}}{|x|}A(k,\theta^{'},\theta)+0\left( \frac{1}{|x|}\right), \\ |x|\rightarrow \infty,
\end{equation}
where   $ A(k,\theta^{'},\theta)  $ is the scattering amplitude and  $ \theta^{'}= \frac{x}{|x|},\theta\in S^{2} $ for  $ k\in \bar{C}^{+}=\{Im k\ge 0\} $ $ \Psi _{0}(k,\theta, x) =e^{ik(\theta, x)}  $
$$\label {eq:3}
A(k,\theta^{'},\theta)=-\frac{1}{4\pi }\int_{R^{3}}q(x)\Psi _{+ } (k,\theta,x)e^{-ik\theta^{'} x}dx.
$$
Solutions ( \ref{eq:1} )- ( \ref{eq:2} ) are obtained by solving the integral equation
$$\label{eq:4}
\Psi_{+ } (k,\theta,x)=\Psi _{0}(k,    
\theta, x)+\int_{R^3}q(y)\frac{e^{+ ik|x-y|}}{|x-y|}\Psi{+ } (k,\theta,y)dy=G(q\Psi_{+ }) 
$$
which is called the Lippman-Schwinger equation.\\
Let inroduce
$$ 
	 \theta,\theta^{'}\in S^{2}, 
 Df=k\int_{S^{2}}A(k,\theta^{'},\theta)f(k,\theta^{'}) d\theta^{'},
$$

Let us also define the solution  $ \Psi _{- } (k,\theta,x)$ for $ k\in \bar{C}^{-}=\{Im k\le 0\} $ as  \[\Psi _{- } (k,\theta,x)=\Psi _{+ } (-k,-\theta,x) \].
As is well known[8] : 
\begin{equation}
\label{eq:eq5}
\Psi _{+ } (k,\theta,x)-\Psi _{- } (k,\theta,x)=
-\frac{k}{4\pi}\int_{S^{2}}A(k,\theta^{'},\theta)\Psi_{-} (k,\theta^{'},x)d\theta^{'},~  k\in R.
\end{equation}
This equation is the key to solving the inverse scattering problem, and was first used by Newton~[8,9] and Somersalo et al.~[10].

\begin{definition}
	The set of measurable functions $\mathbf{R}$ with the norm, defined by $$
	||q||_{\mathbf{R}}=\int_{R^{6}}\frac{q(x)q(y)}{|x-y|^{2}}dxdy<\infty $$
	is recognized as being of Rollnik class.

\end{definition}\label{df:d1}

Equation (\ref{eq:eq5}) is equivalent to the following:
$$
\Psi _{+}=S\Psi _{-},  \label{eq:eq6}
$$
where  $ S $ is a scattering operator with the kernel $S(k,\textit{\l}), 
$ $$ ,~S(k,\textit{\l})=\int_{R^{3}}\Psi _{+}(k,x)\Psi _{-}^{\ast }(\textit{\l},x)dx $$.

The following theorem was stated in  [9]:

\begin{theorem}\label{thm:t1}\textbf{(The energy and momentum conservation laws)}
	Let $q\in \mathbf{R}$. Then, $ SS^{\ast }=I,~S^{\ast }S=I,$ where $ I $ is a unitary operator.
\end{theorem}
\begin{corollary}  $ SS^{\ast }=I,~S^{\ast }S=I,$  yeild
	$$
	A(k,\theta^{'},\theta)-A(k,\theta,\theta^{'})^{\ast }=\frac{ik}{2\pi}\int_{S^{2}}A(k,\theta,\theta^{''})A (k,\theta^{'}, \theta^{''} )^{\ast }d\theta^{''}
	$$
	
\end{corollary}

\begin{theorem}
	\label{Theorem 1.4}
	\textbf{(Birmann--Schwinger estimation).}
	Let $q\in \mathbf{R}$. Then, the number of discrete eigenvalues can be estimated as: 
	$$
	N(q)\leq \frac{1}{(4\pi )^{2}}\int_{R^{3}}\int_{R^{3}}\frac{%
		q(x)q(y)}{|x-y|^{2}}dxdy.
	$$
\end{theorem}
\begin{lemma}\label{lm:l2} 
	Let $
	\left(|q|_{L_{1}(R^3)} +4\pi|q|_{L_{2}(R^3)}   \right)<\alpha<1/2   $. Then, 
$$
	\left\| \Psi_{+}\right\|_{L_{\infty}} \le \frac{\left(|q|_{L_{1}(R^3)} +4\pi|q|_{L_{2}(R^3)}   \right)}{ 1-\left(|q|_{L_{1}(R^3)} +4\pi|q|_{L_{2}(R^3)}   \right)    } < \frac{\alpha}{1-\alpha} 
$$
	$$
	\left\|   \frac{ \partial(\Psi_{+}-\Psi_0)}{\partial k}\right\|_{L_{\infty}}\le  \frac{  	|q|_{L_{1}(R^3)} +4\pi|q|_{L_{2}(R^3)}  }{1-\left(|q|_{L_{1}(R^3)} +4\pi|q|_{L_{2}(R^3)}   \right)    }  < \frac{\alpha}{1-\alpha}     
	$$
\end{lemma}
\begin{proof}
	By Lippman-Schwinger equation's
	$$
	\left| \Psi_{+}-\Psi_0\right| \le  \left| Gq\Psi_{+}\right| ,\,\,\,\,
	$$
	
	$$
	\left| \Psi_{+}-\Psi_0\right|_{L_{\infty}} \le \left| \Psi_{+}-\Psi_0\right|_{L_{\infty}} \left|   Gq\right|+\left|Gq\right| 	$$
	finaly 	
		$$
			\left| \Psi_{+}-\Psi_0\right| \le \frac{\left(|q|_{L_{1}(R^3)} +4\pi|q|_{L_{2}(R^3)}   \right)}{ 1-\left(|q|_{L_{1}(R^3)} +4\pi|q|_{L_{2}(R^3)}   \right)    }  
		$$	
	
		By Lippman-Schwinger equation's
		$$
		\left| \frac{ \partial\left( \Psi_{+}-\Psi_0\right) }{\partial k} \right| \le  \left|\frac{ \partial Gq}{\partial k}   \Psi_{+}\right|+  \left| Gq \frac{ \partial\left( \Psi_{+}-\Psi_0\right) }{\partial k}   \right| +\left| Gq\right|
		$$
		$$\label{eq:eq2}	\left| \frac{ \partial (\Psi_{+}-\Psi_0)}{\partial k} \right|  \le \left(|q|_{L_{1}(R^3)} +4\pi|q|_{L_{2}(R^3)}   \right)  
		$$
		
		$$
		\left\|   \frac{ \partial(\Psi_{+}-\Psi_0)}{\partial k}\right\|_{L_{\infty}}\le  \frac{  	|q|_{L_{1}(R^3)} +4\pi|q|_{L_{2}(R^3)}  }{1-\left(|q|_{L_{1}(R^3)} +4\pi|q|_{L_{2}(R^3)}   \right)    }    
		$$
	proof completes
\end{proof}
Let us introduce the following notations:
$$ 
	Q(k,\theta,\theta^{'})=\int_{R^3}q(x)e^{ik(\theta-\theta^{'})x}dx,K(s)=s,X(x)=x.\,\,\, 	$$
	$$\label{eq:eq2} T_+Q=\int_{-\infty}^{+\infty}\frac{Q(s,\theta,\theta^{'})}{ s-t-i0}ds,\,\,\,T_-Q=\int_{-\infty}^{+\infty}\frac{Q(s,\theta,\theta^{'})}{ s-t+i0}ds .
$$

\begin{lemma}\label{lm:l2} 
	Let $q\in \mathbf{R}\cap  L_{1}(R^3), \left\| q \right\|_{L_1}+4\pi|q|_{L_{2}(R^3)} <\alpha<1/2   $. Then, 
	$$
	\left\|  	A _{+}\right\|_{L_{\infty}}  < \alpha+  \frac{\alpha}{1-\alpha}   .
	$$
	
		$$
		\left\|  \frac{ \partial A _{+}}{\partial k}	\right\|_{L_{\infty}}  < \alpha+  \frac{\alpha}{1-\alpha}.
		$$
	
\end{lemma}
\begin{proof}
	Multiplying Lippman-Schwinger equation on $q(x)\Psi_0(k,\theta,x)$ and after integrating we have
	$$
	A(k,\theta,\theta^{'})=	Q(k,\theta,\theta^{'})+		\int_{R^{3}}q(x)\Psi_0(k,\theta,x)G	q\Psi_{+}dx
	$$
	Estimating latest equation
	$$
	\left| A\right| \le \alpha+ \alpha\frac{\left(|q|_{L_{1}(R^3)} +4\pi|q|_{L_{2}(R^3)}   \right)}{ 1-\left(|q|_{L_{1}(R^3)} +4\pi|q|_{L_{2}(R^3)}   \right)    }.
	$$
Similarly  for $	\left\|  \frac{ \partial A _{+}}{\partial k}	\right\|$	proof completes	
\end{proof}

We define the operators ~$T_{\pm }$, $T$ for ~$f\in W_{2}^{1}(R)$ as follows:
$$ 
T_{+}f=\frac{1}{2\pi i}\lim\limits_{Imz\rightarrow 0}\int\limits_{-\infty}^{\infty }\frac{f({s})}{s-z}ds,~Im~z>0,
T_{-}f=\frac{1}{2\pi i}\lim\limits_{Imz\rightarrow 0}\int\limits_{-\infty
}^{\infty }\frac{f({s})}{s-z}ds,~Im~z<0,
$$
$$
Tf=\frac{1}{2}(T_{+}+T_{-})f.
$$
Consider the Riemann problem of finding a function $\Phi $, that is analytic in the complex plane with a cut along the real axis.Values of
$ \Phi $  on the sides of the cut are denoted as $\Phi_{+}$, $\Phi_{-} $.The following presents the results of [12]:
\begin{lemma}\label{lm:l1}
	$$
	TT= \frac14 I,~TT_+ = \frac12 T_+,~TT_- = - \frac12 T_-, \ T_+ = T+\frac12 I,~T_- = T-\frac12 I,~T_-T_- = -  T_-
	$$ 
\end{lemma}

Denote
$$
\Phi_{+}(k,\theta,x)=\Psi_{+}(k,\theta,x)-\Psi_{0}(k,\theta,x), \,\,\,\Phi_{-}(k,\theta,x)=\Psi_{-}(k,-\theta,x)-\Psi_{0}(k,\theta,x),\,\, 	$$
$$\label{eq:eq2} g(k,\theta,x)=\Phi_{+}(k,\theta,x)-\Phi_{-}(k,\theta,x)
$$


\begin{lemma}\label{lm:l2} 
	Let $q\in \mathbf{R},\,\, N(q)<1 ~g_{+}=g(k,\theta,x), ~g_{-}=g(k,-\theta,x). $. Then, 
	$$
		\Phi _{+}(k,\theta,x)=T_{+ } g_{+}  +e^{ik\theta x},\ \Phi _{-}(k,\theta,x)=T_{- } g_{+}  +e^{ik\theta x}.
	$$ 
\end{lemma}	
\begin{proof}  
The proof of the above follows from the classic results for the Riemann problem.
\end{proof}
\begin{lemma}\label{lm:l2} 
	Let $q\in \mathbf{R},\,\, N(q)<1 ~g_{+}=g(k,\theta,x), ~g_{-}=g(k,-\theta,x) , ) $. Then, 
	$$
	\Psi _{+}(k,\theta,x)=(T_{+ } g_{+}  +e^{ik\theta x}),\ \Psi _{-}(k,\theta,x)=(T_{- } g_{-}  +e^{-ik\theta x}).
	$$ 
\end{lemma}
\begin{proof}
The proof of the above follows from the definitions of $ g , \Phi _{\pm }, \Psi _{\pm } $ .
\end{proof}

\begin{lemma}\label{lm:l6}
	Let $$ \sup\limits_{ k} \left|    \int\limits_{-\infty}^{\infty} \frac{ pA(p,\theta^{'},\theta) }{ 4\pi(p-k+i0 ) }dp  \right|<\alpha,\,\,\int_{S_2}\alpha d\theta<1/2	$$ 
 Then
	$$
	         \prod\limits_{ 0\le j<n }  \int_{S_2}\left|   \int_{-\infty}^{\infty} \frac{ {k_{j}}A(k_{j},\theta^{'}_{k_{j}},\theta_{k_{j}}) }{ 4\pi( k_{j+1}-k_{j}+i0 ) }d{k_{j}} \right |d\theta_{k_{j}} \le 2^{-n}
		\label{lm:psi}
$$
\end{lemma}
\begin{proof}
	
		denote 
		$$
		\alpha_j{}=\left| Vp \int_{-\infty}^{\infty} \frac{ {k_{j}}A(k_{j},\theta^{'}_{k_{j}},\theta_{k_{j}}) }{ 4\pi( k_{j+1}-k_{j}+i0 ) }d{k_{j}}\right| , \,\,\, 
		$$
		therefore
		$$
	        \prod\limits_{ 0\le j<n } \int_{S_2} \left| \int_{-\infty}^{\infty} \frac{ {k_{j}}A(k_{j},\theta^{'}_{k_{j}},\theta_{k_{j}}) }{ 4\pi( k_{j+1}-k_{j}+i0 ) }d{k_{j}}  \right |d\theta_{k_{j}}\le   \prod\limits_{ 0\le j<n } \int_{S_2} \alpha_j{} d\theta_{k_{j}}  <
	   	     2^{-n}.	$$
	     $$\label{eq:eq2}
		$$
		Proof completes.
\end{proof}


			\begin{lemma}\label{lm:TA}
				
			Let 	$$
			\sup\limits_{ k } \int_{S^{2}} \left| T_{-}QK\right|d\theta \le \alpha<\frac{1}{2C}<1,\,\,\,\, \sup\limits_{ k } \int_{S^{2}}\left| T_{-}\tilde{q}K\right|d\theta\le \alpha<\frac{1}{2C}<1	,\,\,\,\, 	$$
			$$\label{eq:eq2}\sup\limits_{ k } \int_{S^{2}}\left| T_{-}Q\tilde{q}K^2\right|d\theta\le \alpha<\frac{1}{2C}<1		$$
			Then
				$$
						\sup\limits_{ k}\int_{S^{2}}\left| 	T_{-}AK\right|d\theta \le\frac{ C\int_{S^{2}}\left|T_{-}QK\right|d\theta}{1-\sup\limits_{ k}\int_{S^{2}}\left| 	T_{-}A\tilde{q}K^2\right|d\theta},\,\,\, 		$$
																$$\label{eq:eq2}	
							\sup\limits_{ k}\left| \int_{S^{2}}	T_{-}A\tilde{q}K^2d\theta\right| \le
							\frac{ C\left|T_{-}\int_{S^{2}}Q\tilde{q}K^2d\theta\right|}{1- \left| T_{-}\int_{S^{2}}\tilde{q}Kd\theta\right|	}
													$$

			\end{lemma}
			\begin{proof}
				By definition amplitude and Lemma 4 
				$$
					A(k,\theta^{'},\theta)=-\frac{1}{4\pi }\int_{R^{3}}q(x)\Psi _{+ } (k,\theta,x)e^{-ik\theta^{'} x}dx =	$$
					$$\label{eq:eq2}-\frac{1}{4\pi }\int_{R^{3}}q(x) \left[ e^{ik\theta^{'} x} +T_+g(k,\theta,\theta^{'}) \right] e^{-ik\theta^{'} x}dx   .
				$$%
				we can rewrite 
			\begin{equation}\label{eq:5}
					A(k,\theta^{'},\theta)=-\frac{1}{4\pi }\int_{R^{3}}q(x) \left[ e^{ik\theta x} +\sum_{n\ge 0}(-T_-D)^n\Psi_0\right]   e^{-ik\theta^{'} x} dx
					\end{equation}
						 Lemma 6 yeild 
				$$
						\sup\limits_{ k}\int_{S^{2}}\left| 	T_{-}AK\right|d\theta\le	\sup\limits_{ k}	\int_{S^{2}}\left| \frac{1}{4\pi }T_{-}QK\right|d\theta  +\frac{\left( \sup\limits_{ k}\int_{S^{2}}\left| 	T_{-}KA\right|d\theta\right) ^2\int_{S^{2}}\left|T_{-} A\tilde{q}K^{2}\right|d\theta }{\left( 1-\sup\limits_{ k}\int_{S^{2}}\left| 	T_{-}KA\right|d\theta\right) ^{2}} 
					$$
					
					Due to the smallness of the terms on the right-hand side, the following estimate follows
						$$
						\sup\limits_{ k}\int_{S^{2}}\left| 	T_{-}AK\right|d\theta\le	2\sup\limits_{ k}	\int_{S^{2}}\left| \frac{1}{4\pi }T_{-}QK\right|d\theta  
						$$
						similarly
						
							$$\label{eq:eq2}		\sup\limits_{ k}\int_{S^{2}}\left| 	T_{-}A\tilde{q}K^2\right|d\theta \le C\int_{S^{2}}\left|T_{-}Q\tilde{q}K^2\right|d\theta+  \int_{S^{2}}\left| T_{-}A\tilde{q}K^2\right|d\theta	\int_{S^{2}} \left| T_{-}\tilde{q}K\right|d\theta	   
							$$	
								$$\label{eq:eq2}	
								\sup\limits_{ k}\int_{S^{2}}\left| 	T_{-}A\tilde{q}K^2\right|d\theta \le
								\frac{ C\int_{S^{2}}\left|T_{-}Q\tilde{q}K^2\right|d\theta}{1- \int_{S^{2}}\left| T_{-}\tilde{q}K\right|d\theta	}
								$$
						$$
						\sup\limits_{ k}\int_{S^{2}}\left| 	T_{-}A\tilde{q}K^2\right|d\theta\le	2\sup\limits_{ k}\int_{S^{2}}	\left| \frac{1}{4\pi }T_{-}Q\tilde{q}K^2\right|d\theta  
						$$

				proof completes	
			\end{proof}
	

	To simplify the writing of the following calculations, we introduce the set defined by
	$$M_{\epsilon}(k)=\left( s|  \epsilon<|s|+|k-s|<\frac{1}{\epsilon}\right) $$
	and function of Heavisid given by 
	$$
	{\Theta}(x) =
	\left\{
	1 ,\,\,\mbox{if } x> 0 ,\,\,\,
	1/2 \,\,\,  \mbox{if } x= 0, \,\,\,\,
	-1 \,\,  \mbox{if } x< 0 \,\,\,
	\right\}.
	$$

	\begin{lemma} \label{lm:l8}
		Let $q,\nabla q\in \cap L_{2}(R^{3})$. Then,
		$$ 
		\int_{R^3} \Theta(A) e^{ik|x|A}q(x)dx  =    \lim\limits_{\epsilon\rightarrow 0}\int_{s\in M_{\epsilon}(k)}\int_{R^3} \frac{e^{is|x|A}}{k-s} q(x)ds   
		$$
		$$ 
		\int_{R^3} \Theta(A) ke^{ik|x|A}q(x)dx  =    \lim\limits_{\epsilon\rightarrow 0}\int_{s\in M_{\epsilon}(k)}\int_{R^3} s\frac{e^{is|x|A}}{k-s} q(x)dxds   
		$$
		
	\end{lemma} 
	\begin{proof}
		The lemma can be proved by conditions of Lemma and Lemma of  Jordan.
	\end{proof}



		\begin{lemma}\label{lm:TkQ}
			
			Let $$ l=2	,\,\,\,I_0= \Psi_0(x,k)|_{ r=r_{0}} 
				$$
			Then
		$$		\left| 	\int_{-\infty}^{+\infty}\int_{S^{2}}\int_{S^{2}}\tilde{q}(k(\theta-\theta'))I_0k^2dkd\theta d\theta'\right| \le \sup\limits_{x\in R^3} \left| q(x)\right| +C_0(\frac{1}{r_{0}} +r_0)\left\| q\right\|_{L_2(R^3)} 
			$$
			
			$$	 	\sup \limits_{\theta\in S^2}\left|\int_{-\infty}^{+\infty} \int_{S^{2}}\int_{S^{2}} QTKQI_0k^2d\theta''d\theta'dk \right|  \le  C_0(\frac{1}{r_{0}} +r_0)\left\| q\right\|^2_{L_2(R^3)} 
			$$	
		\end{lemma}
		\begin{proof}
			By definition Fourier transform

			$$\label{eq:eq2}	
		\int_{-\infty}^{+\infty}\int_{S^{2}}\int_{S^{2}}\tilde{q}(k(\theta-\theta'))I_0k^2dkd\theta d\theta'=
				\int_{-\infty}^{+\infty}\int_{S^{2}}\int_{S^{2}} \int_{0}^{+\infty}q(x)e^{ikx(\theta-\theta')}e^{ix_0k}k^2dkd\theta d\theta' drd\gamma
			$$	
		where $x=r\gamma$
				The lemma of the Jordan completes the first  inequality  proof.
							The second inequality is proved like the first
							
					$$	 \int_{-\infty}^{+\infty} \int_{S^{2}}\int_{S^{2}} QTKQI_0k^2d\theta''d\theta'dk 
					=$$$$
					\int_{-\infty}^{+\infty} \int_{-\infty}^{+\infty} \int_{S^{2}}\int_{S^{2}}\int_{S^{2}} \frac{\left( \tilde{q}(scos(\theta')- scos(\theta''))\tilde{q}(kcos(\theta)-scos(\theta'')\right)s }{k-s}   I_0k^2d\theta'd\theta''d\theta dk ds
					$$				
				Lemma 8 yield			
					$$	\int_{-\infty}^{+\infty}\int_{S^{2}} \int_{S^{2}}\int_{S^{2}} \left( \tilde{q}(kcos(\theta')- kcos(\theta))\tilde{q}(kcos(\theta)-kcos(\theta'')\right) I_0k^3\Theta (cos(\theta''))d\theta'd\theta''d\theta dk-
					$$
							$$	\int_{-\infty}^{+\infty} \int_{S^{2}}\int_{S^{2}} \int_{S^{2}}\left( \tilde{q}(kcos(\theta')- kcos(\theta))\tilde{q}(kcos(\theta)-kcos(\theta'')\right) I_0k^3\Theta (-cos(\theta''))d\theta'd\theta''d\theta dk
							$$
							Integrating $\theta$, $\theta'$, $\theta''$ ,$k$ we obtain the proof of the second inequality of the lemma
								
		\end{proof}

		
		\begin{lemma}\label{lm:TA}
			
			Let 	$$
			\sup\limits_{ k }  \left| T_{-}QK\right| \le \alpha<\frac{1}{2C}<1,\,\,\,\, \sup\limits_{ k } \left| T_{-}\tilde{q}K\right|\le \alpha<\frac{1}{2C}<1	,\,\,\,\, 	$$
			$$\label{eq:eq2}\sup\limits_{ k } \left| T_{-}Q\tilde{q}K^2\right|\le \alpha<\frac{1}{2C}<1	,\,\,\,l=0,1,2		$$
			Then
			
			$$\label{eq:A^0}	
		\left| 	\int_{-\infty}^{+\infty} \int_{S^{2}}\int_{S^{2}} A(k,\theta',\theta)k^ldkd\theta'd\theta \right| \le \left| \int_{-\infty}^{+\infty}\int_{S^{2}}\int_{S^{2}}\tilde{q}(k(\theta-\theta'))k^ldkd\theta' d\theta\right| +
		$$$$
		 C\sup \limits_{\theta\in S^2}\left|\int_{-\infty}^{+\infty} \int_{S^{2}}\int_{S^{2}} QTKAk^ld\theta''d\theta'dk \right| 
												$$
					$$\label{eq:A^0}	
				\left| 	\int_{-\infty}^{+\infty} \int_{S^{2}}\int_{S^{2}} A(k,\theta',\theta)k^2dkd\theta'd\theta \right| \le \sup\limits_{x\in R^3} \left| q\right|  + C_0\left\| q\right\|_{W_2^1(R^3)}\left\| q\right\|_{L_2(R^3)}\left(  \left|\int_{S^{2}}TKAd\theta'' \right| +1  \right) 
				$$

					\end{lemma}
		\begin{proof}

		Definition amplitude , Lemma 3-4 and Lemma of Jordan yeild 
			$$
			\int_{-\infty}^{+\infty}\int_{S^{2}}\int_{S^{2}}	A(k,\theta^{'},\theta)k^ldkd\theta'd\theta =-	\int_{-\infty}^{+\infty}\frac{1}{4\pi }\int_{S^{2}}\int_{S^{2}}\int_{R^{3}}q(x)\Psi _{+ } (k,\theta,x)e^{-ik\theta^{'} x}k^ldxdkd\theta' =	$$
			$$\label{eq:eq2}-\frac{1}{4\pi }\int_{S^{2}}\int_{S^{2}}\int_{R^{3}}q(x) \left[ e^{ik\theta x} +\sum_{n\ge  1}(-T_-D)^n\Psi_0\right]  e^{-ik\theta^{'} x}k^ld\theta' dxdk =
			$$$$
			\int_{-\infty}^{+\infty}\int_{S^{2}}\int_{S^{2}}\tilde{q}(k(\theta-\theta'))k^ldkd\theta'd\theta+ \sum_{n\ge  1}W_n 
			$$%
			$$
			W_1= \int_{R^{3}}\int_{-\infty}^{+\infty}\int_{S^{2}}\int_{S^{2}}\frac{sA(s,\theta^{''},\theta)e^{-ik\theta^{'} x}q(x)e^{i s\theta'' x }}{k-s}k^l dkdxdsd\theta'd\theta''
			$$
		
				$$
			\left| 	W_1\right| \le  C\sup \limits_{\theta\in S^2}\left|\int_{-\infty}^{+\infty} \int_{S^{2}}\int_{S^{2}} QTKAk^ld\theta''d\theta'dk \right|  
				$$

				similarly 
				
					$$
					\left| 	W_n\right| \le C\sup \limits_{\theta\in S^2}\left|\int_{-\infty}^{+\infty} \int_{S^{2}}\int_{S^{2}} QTKAk^ld\theta''d\theta'dk \right| \left|\int_{S^{2}}TKAd\theta'' \right|^n
					$$
				Finally	
				
					$$
					\left| 	\int_{-\infty}^{+\infty} \int_{S^{2}}\int_{S^{2}} A(k,\theta',\theta)dkd\theta'd \theta \right| \le \left| \int_{-\infty}^{+\infty}\int_{S^{2}}\int_{S^{2}}\tilde{q}(k(\theta-\theta'))dkd\theta d\theta'\right|+
					$$$$
					 C_0\left\| q\right\|^2_{L_2(R^3)}\left(  \left|\int_{S^{2}}TKAd\theta'' \right| +1  \right) 
					$$
				
					$$
					\left| 	\int_{-\infty}^{+\infty} \int_{S^{2}}\int_{S^{2}} A(k,\theta',\theta)k^2dkd\theta'\right| \le \sup\limits_{x\in R^3} \left| q\right|  + C_0\left\| q\right\|^2_{L_2(R^3)}\left(  \left|\int_{S^{2}}TKAd\theta'' \right| +1  \right) 
					$$

		proof completes	
	\end{proof}

	\begin{lemma}\label{lm:l6}
		Let $$  \sup\limits_{ k}  \int_{S^{2}}\left|    \int\limits_{-\infty}^{\infty} \frac{pA(p,\theta^{'},\theta) }{ 4\pi(p-k+i0 ) }dp  \right|d\theta<\alpha<1/2,\,\,\, \sup\limits_{ k} \left| pA(p,\theta^{'},\theta)\right|<\alpha<1/2 
		$$  Then
		$$
		|T_{- }D\Psi _{0} |< \frac{\alpha }{1-\alpha},\,\,\,\, |T_{+ }D\Psi _{0} |< \frac{\alpha }{1-\alpha}, \,\,\,\,\,\,\,\,\,
		|D\Psi _{0} |< \frac{\alpha }{1-\alpha} 	$$
		$$\label{eq:eq2}
		\,\, 	T_{- }g_{-}=( I-T_{- }D )^{-1 }T_{- }D\Psi _{0},
		\,\,\,\,\,\,\,\,\,\,\,\,\,\Psi _{-}=( I-T_{- }D )^{-1 }T_{- }D\Psi _{0} +\Psi _{0},           
		\label{lm:psi}
		$$
		q satisfies  the following inequalities:
		$$
		\sup\limits_{x\in R^3}|q(x)|\le \left| \int_{S^{2}}TKQ d\theta\right|
		C_0\left( \left\| q\right\|^2_{L_2(R^3)}+1\right) +C_0\left\| q\right\|_{L_2(R^3)}
		$$
	\end{lemma}
	\begin{proof}
		Using equation 
		$$
		\Psi _{+ } (k,\theta,x)-\Psi _{- } (k,\theta,x)=
		-\frac{k}{4\pi}\int_{S^{2}}A(k,\theta^{'},\theta)\Psi_{-} (k,\theta^{'},x)d\theta^{'},~  k\in R.
		$$%
		we can rewrite 
		$$
		T_{+}g_{+}-T_{- }g_{-}=D (T_{-}g_{-}+\Psi_{0})
		$$
		Applying the operator $T_{-}$  last   equation we have 
		$$
		T_{- }g_{-}=T_{- }D (T_{-}g_{-}+\Psi_{0})
		$$
		$$
		(I  - T_{- }D    ) T_{- }g_{-}=T_{- }D\Psi_{0},\,\,\,
		T_{- }g_{-}=   \sum_{n\ge0}      \left( -T_{- }D \right)^{n} \Psi_{0}
		$$
		Estimating the terms of the series, we obtain using Lemma 4
		$$
		|(T_{- }D)^{n}\Psi_{0}| \le 
		\sum_{n\ge 0 }
		\left|  \int_{-\infty}^{\infty}.... \int_{-\infty}^{\infty} \Psi_{0}  \prod\limits_{ 0\le j<n }\frac{ \int_{S^{2}}{k_{j}}A(k_{j},\theta^{'}_{k_{j}},\theta_{k_{j}}) d\theta^{'}_{k_{j}}}{ 4\pi(k_{j+1})-k_{j}+i0 ) }   dk_{1}...d_{k_{n}}\right| \le 	$$
		$$\label{eq:eq2}\\ \sum_{n> 0 } 2^n \alpha^{n} =\frac{2\alpha}{1-2\alpha}  
		$$
			Denote $$\Lambda=  \frac{\partial }{\partial k}, r=\sqrt{x_1^2+x_2^2+x_3^2} $$ 
			
			we have
			$$
			\Lambda\int_{S^{2}}\Psi_0d\theta =\Lambda  \frac{sin( kr)}{ikr}=  
	 \frac{cos(kr)}{ik}-\frac{sin( kr)}{ik^2r} 
			$$
			$$
			\Lambda\int_{S^{2}}H_0\Psi_0d\theta =\Lambda k^2 \frac{sin( kr)}{ikr}=  
			k\frac{cos(kr)}{i}+\frac{sin( kr)}{ik^2r} 
			$$
			$$
			\left| \Lambda \int_{S^{2}}\Psi d\theta	 \right|=\left|\Lambda\int_{S^{2}}\Psi_0d\theta +\Lambda\int_{S^{2}}\sum_{n\ge 0}\left( -T_{- }D \right)^{n} \Psi_{0} d\theta\right|  > (\frac{1}{k} -  \frac{\alpha}{1-\alpha}), as\,\, kr=\pi.
			$$
		
							and
					$$
					\Lambda\frac{1}{k-t}=
					-   \frac{1}{(k-t)^2} 
					$$

	Equation ( \ref{eq:1} )  yield 
		$$
		q= \frac{\Lambda\left(  H_0\int_{S^{2}}\Psi d\theta+k^2\int_{S^{2}}\Psi d\theta\right)  }{\Lambda\int_{S^{2}}\Psi d\theta}= $$$$
		 \frac{2k\int_{S^{2}}T_{-}g_{-}d\theta+k^2\int_{S^{2}}\Lambda T_{-}g_{-}d\theta+  H_0\Lambda \int_{S^{2}}T_{-}g_{-}d\theta}{\Lambda\int_{S^{2}}\Psi d\theta} = $$
		
				$$
							\frac{2k\int_{S^{2}}T_{-}g_{-}d\theta+\Lambda\int_{S^{2}}\sum_{n\ge 1}      \left( -T_{- }D \right)^{n}(K^2-k^2) \Psi_{0} d\theta}{\Lambda\int_{S^{2}}\Psi d\theta}=$$$$
									\frac{W_0+\sum_{n\ge 1} \int_{S^{2}}W_{n}}{\Lambda\int_{S^{2}}\Psi d\theta}
				$$

			Denote $$Z(k,s)=s+2k +\frac{2k^2}{k-s}$$	then 
				$$
				\left|  W_1\right| \le 	\left|\int_{-\infty}^{+\infty}\int_{S^{2}} \int_{S^{2}} A(s,\theta,\theta')s \frac{s^2-k^2}{(k-s)^2}  \Psi_0  sin(\theta)ds d\theta\right|_{k=k_0} \le
				$$$$
				 \left| \int_{-\infty}^{+\infty}\int_{S^{2}}\int_{S^{2}}Z(k,)\tilde{q}(k(\theta-\theta'))\Psi_0dkd\theta\right|+
									C_0\left|\int_{S^{2}} TKQd\theta\right|
					$$
					
		For calculating $W_{n}$, as $n \ge 1$ take simple transformation
			
			$$
			\frac{s_{n}^3}{s_{n}-s_{n-1}}=	\frac{s_{n}^3-s_{n}^2s_{n-1}}{s_{n}-s_{n-1}} +\frac{s_{n}^2s_{n-1}}{s_{n}-s_{n-1}}=s_{n}^2+\frac{s_{n}^2s_{n-1}}{s_{n}-s_{n-1}}=
			$$
			
			\begin{equation} \label{eq:6}
			s_{n}^2+\frac{s_{n}^2s_{n-1}-s_{n}s_{n-1}^2}{s_{n}-s_{n-1}} +\frac{s_{n}s_{n-1}^2}{s_{n}-s_{n-1}}=	s_{n}^2+s_{n}s_{n-1}+\frac{s_{n}s_{n-1}^2}{s_{n}-s_{n-1}}
			\end{equation}
				$$
		\frac{As_{n}^3}{s_{n}-s_{n-1}} =As_{n}^2+As_{n}s_{n-1}+    \frac{As_{n}s^2_{n-1}}{s_{n}-s_{n-1}}=V_1+V_2+V_3 
				$$
		Using Lemma \ref{lm:TA},   for estimating $V_1$ , $V_2$ . For $ V_3 $ take again simple transformation for $s^3_{n-1} $ which will appear in the integration over  $ s_{n-1}$, finally   we get 
		
		$$
			|q(x)|_{r=r_0}= \left| \frac{\Lambda\left(  H_0\int_{S^{2}}\Psi d\theta+k^2\int_{S^{2}}\Psi d\theta\right)  }{\Lambda\int_{S^{2}}\Psi d\theta}\right| _{k=k_0,r= \frac{\pi}{k_0}}\le $$$$
			\frac{\left| \int_{-\infty}^{+\infty}\int_{S^{2}}\int_{S^{2}}Z(k,)\tilde{q}(k(\theta-\theta'))\Psi_0dkd\theta d\theta'\right|+
				C_0\left|\int_{S^{2}} TKQd\theta\right|}{(\frac{1}{k_0}-\frac{\alpha}{(1-\alpha)} )}   + 
			$$
		$$
			$$	
Finally we get
		
			$$
			|q(x)|_{r=r_0}\le \sup\limits_{x\in R^3}|q(x)|\alpha+
		 C_0\left\| q\right\|^2_{L_2(R^3)}+C_0\left\| q\right\|_{L_2(R^3)}+\left| \int_{S^{2}}TKQ d\theta\right|
			$$
		
		The invariance of the Schrödinger equations with respect to translations and the arbitrariness $r_0$	yield
			$$
			\sup\limits_{x\in R^3}|q(x)|\le \left| \int_{S^{2}}TKQ d\theta\right|
			C_0\left( \left\| q\right\|^2_{L_2(R^3)}+1\right) +C_0\left\| q\right\|_{L_2(R^3)}
			$$
			\end{proof}


\section{Conclusions for the three-dimensional inverse scattering problem }
This study has shown once again the outstanding properties of the scattering operator, which , in combination with the analytical properties of the wave function,allow to obtain an almost- explicit formulas for the potential to be obtained  from the scattering amplitude. Furthermore, this appro. The estimations follow from this reach overcomes the problem of over-determination, resulting from the fact that the potential is a function of three variables, whereas the amplitude is a function of five variables. We have shown that it is sufficient to average the scattering amplitude to eliminate the two extra variables.
\section{Cauchy problem for the Navier--Stokes equation}
Numerous studies of the Navier-Stokes equations have been devoted to the problem of the smoothness of its solutions. A good overview of these studies is given in [13]-[17]. The spatial differentiability of the solutions is an important factor, this controls their evolution. Obviously, differentiable solutions do not provide an effective description of turbulence. Nevertheless, the global solvability and differentiability of the solutions has not been proven, and therefore the problem of describing turbulence remains open. It is interesting to study the properties of the Fourier transform of solutions of the Navier-Stokes equations. Of particular interest is how they can be used in the description of turbulence, and whether they are differentiable. The differentiability of such Fourier transforms appears to be related to the appearance or disappearance of resonance, as this implies the absence of large energy flows from small to large harmonics, which in turn precludes the appearance of turbulence.
Thus, obtaining uniform global estimations of the Fourier transform of solutions of the Navier-Stokes equations means that the principle modeling of complex flows and related calculations will be based on the Fourier transform method. The authors are continuing to research these issues in relation to a numerical weather prediction model; this paper provides a theoretical justification for this approach. Consider the Cauchy problem for the Navier-Stokes equations:
\begin{equation}
\label{1}
  \frac{\partial\vec{v}}{\partial t}-\nu \Delta \vec{v}+(\vec{v},\nabla \vec{v})=-\nabla
p+\vec{f}(x,t),~div~\vec{v}=0,  
\end{equation}
\begin{equation}\label{2}
\vec{v}|_{t=0}=\vec{v}_{0}(x)  
\end{equation}
in the domain $Q_{T}=R^{3}\times (0,T)$,where :
\begin{equation}
div\;\vec{v}_{0}=0.  \label{3}
\end{equation}
The problem defined by  (\ref{1}), (\ref{2}), (\ref{3}) has at least one weak solution $ (\vec{v}, p) $ in the so-called Leray--Hopf class [16].
The following  results have been  proved  [15]:
\begin{theorem}
	\textbf{} If
	$$
	\vec{v}_{0}\in W_{2}^{1}(R^{3}),\;\;\vec{f}(x,t)\in L_{2}(Q_{T}),
	$$%
	there is a single generalized solution of (\ref{1}), (\ref{2}), (\ref{3}) in the domain $Q_{T_{1}}$, $T_{1}\in \lbrack 0,T]$, satisfying the following conditions:
	$$
	\vec{v},\nabla ^{2}\vec{v},\ \ \ \nabla p\in L_{2}(Q_{T}).
	$$
	\label{thm1}
\end{theorem}
Note that $T_{1}$ depends on $\vec{v}_{0}$ and $\vec{f}(x,t)$.
\begin{lemma}\label{lm:l8}
	Let $\vec{v_0} \in W_{2}^{2}(R^{3}),\vec{f} \in L_2(Q_T)  $,
	Then the solution of (\ref{1}), (\ref{2}), (\ref{3}) satisfies  the following inequalities:
	$$
	\sup\limits_{0\leq t\leq
		T}||\vec{v}||_{L_{2}(R^{3})}^{2}+\nu\int\limits_{0}^{t}||\nabla \vec{v}||_{L_{2}(R^{3})}^{2}d\tau  \leq \ ||\vec{v}_{0}||_{L_{2}(R^{3})}^{2}+||\vec{f}||_{L_{2}(Q_{T})}.	$$
	$$\label{eq:eq2}
		\sup\limits_{0\leq t\leq
			T}||\vec{\nabla v}||_{L_{2}(R^{3})}^{2}+\nu\int\limits_{0}^{t}||H_{0} \vec{v}||_{L_{2}(R^{3})}^{2}d\tau  \leq 	$$
		$$\label{eq:eq2} || \nabla\vec{v}_{0}||_{L_{2}(R^{3})}^{2}+||\vec{f}||_{L_{2}(Q_{T})} +\int_{0}^{t}||(\vec{v},\nabla \vec{v})||_{L_{2}(R^{3})}||H_{0} \vec{v}||_{L_{2}(R^{3})}	$$
		$$\label{eq:eq2}
	\nu\int\limits_{0}^{t}||H_{0} \vec{v}||_{L_{2}(R^{3})}^{2}d\tau \le C+ \frac{1}{\nu}\int_{0}^{t}||(\vec{v},\nabla \vec{v})||^{2}_{L_{2}(R^{3})}dt	
	$$
\end{lemma}
\begin{lemma}
Let $\vec{v_0} \in W_{2}^{2}(R^{3}), \vec{\tilde{v_0}} \in W_{2}^{2}(R^{3}),\vec{f} \in L_2(Q_T) $,
Then the solution of (\ref{1}), (\ref{2}), (\ref{3}) satisfies  the following inequalities:
	$$
	\widetilde{\vec{v}}=\widetilde{\vec{v}}_{0}+
	\int\limits_{0}^{t}e^{-\nu k^{2}|(t-\tau )}(\widetilde{[(\vec{v},\nabla )\vec{v}]}+\widetilde{\vec{F}})
	d\tau ,
	\label{eqno8.4}
	$$
	where $\vec{F}=-\nabla p+\vec{f}$.
\end{lemma}
\begin{proof}
	This follows from the definition of the Fourier transform and the theory of linear differential equations.
\end{proof}


Let us introduce operators $F_{k},F_{kk\prime}$, as $$F_{k}f=\int_{R^{3}} e^{i(k,x)} f(x)dx,\,\,\,F_{kk\prime}f=\int_{R^{3}} e^{i(k,x)-i(x,k^{\prime})} f(x)dx $$
$$\vec{\tilde{v}} (k)=F_{k}\vec{v},\,\, \vec{V} (k,k^{\prime})=F_{kk\prime}\vec{v}= \int_{R^{3}} e^{i(k,x)-i(x,k^{\prime})}\vec{v}dx$$

\begin{lemma}\label{lm:19}
	Let $\vec{v_0} \in W_{2}^{2}(R^{3}), \vec{f}\in L_2(Q_T)$,$ \left| TKV_0\right| +\left| TKV_0\right| +\left| TK^2V_0\vec{\tilde{v_0}}\right| <C$.Then,
	the solution of (\ref{1}), (\ref{2}), (\ref{3}) in Theorem \ref{thm1} satisfies  the following inequalities:
	
	$$
	|\tilde{v} (k)|<C,\,\,
	,\,\,
	$$
	$$\label{eq:eq2}
	|TK\tilde{v} (k)|<C_0||v||_{L_2(R^3)}+   \frac{C_0t}{\sqrt{\nu}}||\nabla v||_{L_2(R^3)}||v||_{L_2(R^3)}
	$$,
\end{lemma}
\begin{proof}
	This follows from 
	
	$$
	\vec{\dot v}=
	-(\vec{v}\nabla )\vec{v}+(\nu  \vec{v} + \nabla p) + F ,\\
	$$	
	$$
	\vec{\tilde{v}} = \vec{\tilde{v}}_0+
	\int_{0}^{t} e^{-\nu k^2(t-\tau)}F_{k}\left( -\ (\vec{v},\nabla )\vec{v})+	  \nabla p + 	F \right)d\tau .
	$$	
	from last equation we have

	$$
	|\vec{v} |\le |\vec{v} _0|+C_{T}
	$$

	Denote $$\beta=\sqrt{\nu(t-\tau)},\,\,\,a=\theta x$$
formula  121 (23) from [11] as n=0: yield	
	$$
	\left| 	TK\vec{v} \right| <\left| k e^{-\beta^2k^2}\right|  + \sqrt{\pi}\beta^{-1}e^{- \frac{a^2}{8\beta^2}}D_{0}(\frac{a}{\sqrt{2}\beta }) 		$$
	$$\label{eq:eq2}
	\left| TK\vec{v} \right| \le	\left| TK\vec{v}_0\right| +	$$
	$$\label{eq:eq2}	
	\left| TK\int_{0}^{t} e^{-\nu k^2(t-\tau)}F_{k}\left( -	 (\vec{v},\nabla )\vec{v}]+ \nabla p + 	F\right)dk\right| \le	$$
	$$\label{eq:eq2}	
	\left| TK\vec{v}_0\right| +
	\int_{0}^{t}\left| k e^{-\beta^2k^2}\right|  +	\left|  \sqrt{\pi}\beta^{-1}e^{- \frac{a^2}{8\beta^2}}D_{0}(\frac{a}{\sqrt{2}\beta }) \right|||\nabla \vec{v}||_{L_{2}(R^3)}dt \le
	$$$$
	C_0||v||_{L_2(R^3)}+   \frac{C_0t}{\sqrt{\nu}}||\nabla v||_{L_2(R^3)}||v||_{L_2(R^3)}	
	$$

\end{proof}

		\begin{lemma}\label{lm:19}
	Let $\vec{v_0} \in W_{2}^{2}(R^{3}), \vec{f}\in L_2(Q_T)$,$ \left| TKV_0\right| +\left| TKV_0\right| +\left| TK^2V_0\vec{\tilde{v_0}}\right|+
	\,\,$.Then,
	the solution of (\ref{1}), (\ref{2}), (\ref{3}) in Theorem \ref{thm1} satisfies  the following inequalities:
	$$
		 |\vec{V}  (k,k^{\prime})|<C,\,\,
	 k|\vec{V}  (k,k^{\prime})|<\frac{C}{ \sqrt{(1-cos(\theta))}}	,\,\,	$$
		$$\label{eq:eq2}	 |T\vec{V} K|<C_0||v||_{L_2(R^3)}+   \frac{C_0t}{\sqrt{\nu(1-cos(\theta))}}||\nabla v||_{L_2(R^3)}||v||_{L_2(R^3)}	$$
	
\end{lemma}
\begin{proof}
	This follows from 
	$$
		 \vec{\dot V}=
		-F_{kk\prime}[(\vec{v},\nabla )\vec{v}]+F_{kk\prime} (\nu\Delta \vec{v} +\nabla p) + F_{kk\prime}{F} 
	$$
	
	After the transformations we obtain

	$$
		\vec{\dot V}=
		-F_{kk\prime}[(\vec{v}\nabla )\vec{v}]+(\nu_{k} F_{kk\prime}  \vec{v} +F_{kk\prime} \nabla p) + F_{kk\prime}{F} ,\\
	$$	
	$$
		\vec{V} = \vec{V}_0+
		\int_{0}^{t} e^{-\nu k^2(1-cos(\theta))(t-\tau)}\left( -	F_{kk\prime}[(\vec{v},\nabla )\vec{v}]+	 F_{kk\prime} \nabla p + 	F_{kk\prime}{F} \right).
	$$	
	from last equation we have

	$$
		|\vec{V} |\le |\vec{V} _0|+C_0\int_{0}^{t}||\nabla v||_{L_2(R^3)}||v||_{L_2(R^3)}	d\tau
	$$

	Denote $\beta=\sqrt{(1-cos(\theta))(t-\tau)\nu}$,$a=(\theta-\theta')x$ 	formula  121 (23) from [11] as n=0: yield	
		$$
			\left| 	TK\vec{V} \right| <\left| k e^{-\beta^2k^2}\right|  + \sqrt{\pi}\beta^{-1}e^{- \frac{a^2}{8\beta^2}}D_{0}(\frac{a}{\sqrt{2}\beta }) 		$$
			$$\label{eq:eq2}
			\left| TK\vec{V} \right| \le	\left| TK\vec{V}_0\right| +	$$
			$$\label{eq:eq2}	
			\left| TK\int_{0}^{t} e^{-\nu k^2(1-cos(\theta))(t-\tau)}\left( -	F_{kk\prime} (\vec{v},\nabla )\vec{v}]+	 F_{kk\prime} \nabla p + 	F_{kk\prime}{F} \right)dk\right| \le	$$
			$$\label{eq:eq2}	
					\left| TK\vec{V}_0\right| +
				\int_{0}^{t}\left| k e^{-\beta^2k^2}\right|  +	\left|  \sqrt{\pi}\beta^{-1}e^{- \frac{a^2}{8\beta^2}}D_{0}(\frac{a}{\sqrt{2}\beta }) \right|||\nabla \vec{v}||_{L_{2}(R^3)}|| \vec{v}||_{L_{2}(R^3)}dt <
				$$$$				
				C_0||v||_{L_2(R^3)}+   \frac{C_0t}{\sqrt{\nu(1-cos(\theta)) }}||\nabla v||_{L_2(R^3)}||v||_{L_2(R^3)}
				$$

\end{proof}


\begin{theorem}
	\textbf{\label{Theorem 8.6}} 
Let $\vec{v_0} \in W_{2}^{2}(R^{3}), \vec{f}\in L_2(Q_T),\vec{\tilde{f}}\in W_2^{2,1}(Q_T)$,$ \left| TKV_0\right| +\left| TKV_0\right| +\left| TK^2V_0\vec{\tilde{v_0}}\right| <C,\,\,\int_{0}^{\infty}||  H_0\vec{f}||_{L_{2}(R^{3})}dt<\infty $.Then,
	the solution of (\ref{1}), (\ref{2}), (\ref{3}) in Theorem \ref{thm1} satisfies  the following inequalities:
	$$
		\sup\limits_{x\in R^{3}}||\vec{v}(x)||<C
	$$
		$$
		\left\vert \left\vert\nabla 
		\vec{v}\right\vert \right\vert
		_{L_{2}(R^{3})}+
		\nu\int\limits_{0}^{t}\int\limits_{R^{3}}|H_{0} \vec{v}
		|^{2}dxd\tau \leq const 
		$$
\end{theorem}
\begin{proof}
	
	 Consider the Cauchy problem for the Navier-Stokes equations:
	 \begin{equation}
	 \label{1a}
	 \frac{\partial\vec{v}}{\partial t}-\nu \Delta \vec{v}+(\vec{v},\nabla \vec{v})=-\nabla
	 p+\vec{f}(x,t),~div~\vec{v}=0,  
	 \end{equation}
	 \begin{equation}\label{2a}
	 \vec{v}|_{t=0}=\vec{v}_{0}(x)  
	 \end{equation}
	 in the domain $Q_{T}=R^{3}\times (0,T)$,where :
	 \begin{equation}
	 div\;\vec{v}_{0}=0.  \label{3a}
	 \end{equation}
	We perform the following transformations:
	$$
	\vec{u_{\epsilon}}=\epsilon\vec{v}, p_{\epsilon}=p\epsilon,\,\,f_{\epsilon}=f{\epsilon^2},\,\,\nu_{\epsilon}=\epsilon\nu, s=\frac{t}{\epsilon}
	$$
	then 
	\begin{equation}
	\label{1a}
	\frac{\partial\vec{u_{\epsilon}}}{\partial s}-\nu \Delta \vec{u_{\epsilon}}+(\vec{u_{\epsilon}},\nabla \vec{u_{\epsilon}})=-\nabla_{\epsilon}
	p_{\epsilon}+\vec{f_{\epsilon}}(x,t),~div~\vec{u_{\epsilon}}=0,  
	\end{equation}
	\begin{equation}\label{2a}
	\vec{u_{\epsilon}}|_{t=0}=\vec{u_{\epsilon}}_{0}(x)  
	\end{equation}
	in the domain $Q_{T}=R^{3}\times (0, T_{\epsilon})$,where :
	\begin{equation}
	div\;\vec{u_{\epsilon}}|_{t=0}=0.  \label{3a}
	\end{equation}
	Let us return for convenience to the notation $v_{i}=u_{\epsilon_{i}}$	,
	using equation  for each   $v_{i}=u_{\epsilon_{i}} $	
	
	$$
	-\Delta_{x}\Psi  +v_{i}\Psi  =k^{2}\Psi, ~k\in C  
	\label{eq:se}
	$$
		Using Lemmas 2-25  we get estimates for 
	$$ 
	A_{i},\vec{V}_{i}, TA_{i},T\vec{V}_{i}, kA_{i},k\vec{V}_{i}, TKA_{i},TK\vec{V}_{i},\,TK\tilde{v_{i}},\,TK^2V\tilde{v_{i}}, $$
	Last estimations yield  representation 	
	$$ 
	q= \frac{\Lambda\left(  H_0\int_{S^{2}}\Psi d\theta+k^2\int_{S^{2}}\Psi d\theta\right)  }{\Lambda\int_{S^{2}}\Psi d\theta}|_{r=\frac{\pi}{k_0} ,k=k_0}
	$$
		and   Lemma \ref{lm:l6}	implies	
		
			$$\label{eq:eq2}
			\left\vert \left\vert\nabla 
			\vec{v}\right\vert \right\vert^2
			_{L_{2}(R^{3})}+\nu_{\epsilon} \int\limits_{0}^{t}||H_{0} \vec{v}||_{L_{2}(R^{3})}^{2}d\tau \le \int_{0}^{\infty}|| ( \vec{v})||_{L_{2}(R^{3})}|| ||  H_0\vec{f}||_{L_{2}(R^{3})}dt  + $$ $$
	\left\vert \left\vert\nabla 
	\vec{v_{0}}\right\vert \right\vert^2
	_{L_{2}(R^{3})}
	+ \frac{C_0}{\nu_{\epsilon}}\int_{0}^{t} \left(\frac{C_1}{\nu_{\epsilon}} ||(\nabla \vec{v})||^{2}_{L_{2}(R^{3})}||( \vec{v})||^{2}_{L_{2}(R^{3})} +|| \vec{v}||^{2}_{L_{2}(R^{3})}\right) ||(\nabla \vec{v})||^{2}_{L_{2}(R^{3})}ds
			$$
	Denote $$
	\alpha(s)=\frac{C_0}{\nu_{\epsilon}} \left(\frac{C_1}{\nu_{\epsilon}} ||(\nabla \vec{v})||^{2}_{L_{2}(R^{3})}||( \vec{v})||^{2}_{L_{2}(R^{3})} +|| \vec{v}||^{2}_{L_{2}(R^{3})}\right) 
	$$	
	
	$$\int_{0}^{ \frac{T}{T\epsilon\nu}}\alpha(s) ds \le \int_{0}^{\frac{1}{\nu\epsilon} }\frac{C_0}{\nu_{\epsilon}} \left(\frac{C_1}{\nu_{\epsilon}} ||(\nabla \vec{v})||^{2}_{L_{2}(R^{3})}||( \vec{v})||^{2}_{L_{2}(R^{3})} +|| \vec{v}||^{2}_{L_{2}(R^{3})}\right)ds  \le
	$$
	$$
\frac{C_0C_1}{\nu^3_{\epsilon}}\sup\limits_{ t} || ( \vec{v})||^{2}_{L_{2}(R^{3})}\int_{0}^{\infty}\nu_{\epsilon}||(\nabla \vec{v})||^{2}_{L_{2}(R^{3})}||ds +\frac{C_0}{\nu_{\epsilon}}\sup\limits_{ t} || ( \vec{v})||^{2}_{L_{2}(R^{3})}\le
	$$
		$$
	\frac{C_0  \epsilon^4}{\epsilon\nu^3_{\epsilon}}	+
	\frac{C_0\epsilon^2 \frac{\nu}{\epsilon}}{\nu_{\epsilon}}\le 2C_0	
		$$
	as $\epsilon=\nu \epsilon_0 $
	The Gronwall-Bellman  Lemma  yield  
		$$
		\left\vert \left\vert\nabla 
		\vec{v}\right\vert \right\vert^2
		_{L_{2}(R^{3})}+
		\nu_{\epsilon}\int\limits_{0}^{t}\int\limits_{R^{3}}|H_{0} \vec{v}
		|^{2}dxd\tau \leq  	\left\vert \left\vert\nabla 
		\vec{v_0}\right\vert \right\vert^2_ {L_{2}(R^{3})} e^{2C_0}+$$$$
		e^{2C_0}\int_{0}^{\infty}|| ( \vec{v})||_{L_{2}(R^{3})}|| ||  H_0\vec{f}||_{L_{2}(R^{3})}dt
		$$
\end{proof}
Theorem \ref{Theorem 8.6}  \,\,\,asserts the global solvability and uniqueness of the Cauchy problem for the Navier-Stokes equations.



\section{Conclusions}
Uniform global estimations of the Fourier transform of solutions of the Navier--Stokes equations indicate that the principle  modeling of complex flows and related calculations can be based on the Fourier transform method. In terms of the Fourier transform, under both smooth initial conditions and right-hand sides, no appear exacerbations appear in the speed and pressure modes.A loss of smoothness in terms of the Fourier transform can only be expected  in the case of singular initial conditions, or of unlimited forces in  $ L_ {2} (Q_ {T}) $.
The theory developed by us is supported by numerical calculations carried out in the works [18-20]
Where the dependence of the smoothness of the solution on the oscillations of the system is clearly deduced.
\section{Acknowledgements}
	The author thanks the National Engineering Academy of the Republic of Kazakhstan, in particular, Academician NAS RK B.Zhumagulov for constant attention and support.	
	Moreover, the author thanks the Mathematics seminar at the Kazakhstan branch of the Moscow State University for attention and valuable comments, as well as Professor M. Otelbaev.The author is especially grateful to Professor Plotnikov P.I for a thorough analysis of the work and detailed recommendations that have significantly improved the paper.

\end{document}